\documentclass[aps,prl,twocolumn,groupedaddress]{revtex4-2}
\usepackage{graphics} 
\begin{document}
\title{Cauchy universality and random billiards}
\author{Roberto Artuso$^{1,2}$}
\email{roberto.artuso@uninsubria.it}
\author{Dario Javier Zamora$^{1,3}$}
\email{dariojavier.zamora@uninsubria.it}
\affiliation{$^{1}$Center for Nonlinear and Complex Systems and
  Dipartimento di Scienza e Alta Tecnologia, Via Valleggio 11, 22100
  Como (Italy)} 
\affiliation{$^{2}$I.N.F.N., Sezione di Milano, Via Celoria 16, 20133
  Milano (Italy)} 
\affiliation{$^{3}$Instituto de Fisica del Noroeste Argentino, CONICET and
Universidad Nacional de Tucuman, Av. Independencia 1800, Tucuman, CP 4000
(Argentina)}
\date{\today}
\begin{abstract}
Motion in bounded domains represents a paradigm in several settings: from billiard dynamics, to random walks in a finite lattice, with applications to relevant physical,
ecological and biological problems. A remarkable universal property, involving the average of return times to the boundary, has been theoretically proposed, and experimentally verified in quite different contexts. We discuss here mechanisms that lead to violations of universality, induced by boundary effects. We suggest that our analysis should be relevant where non homogeneity appears in the stationary probability distribution in bounded domain. 
\end{abstract}
\maketitle
What do neutrons and ants have in common? A possible answer, generalizing a geometric result established by Cauchy~\cite{Cau}, is that their average residence time $\langle \tau \rangle$ in a d-dimensional bounded domain $\Omega$ does not depend on the nature of their dynamics, nor on the shape of the bounded region: it is proportional to the ratio between the volume $V_{\Omega}$ and the surface $\Sigma_{\Omega}$ of $\Omega$: $\langle \tau \rangle =\eta_d V_{\Omega}/\Sigma_{\Omega}=\tau_C$, where $\eta_d$ is a numerical factor depending only on space dimensionality d (in the case we will consider in detail d=2 and $\eta_2=\pi$).\\ 
More precisely, Cauchy theorem concerns the case where $\langle \tau \rangle$ is the mean length of randomly distributed chords intersecting a convex body: geometric generalizations of this results have been discussed since then (see \cite{San,Maz08,Cab}). From a physical perspective, Cauchy theorem plays an important role in evaluating the residence time of neutrons (assuming they have a constant speed $v_0$) in a bounded region, provided their density is homogeneous and isotropic, and the medium is non scattering \cite{PAM,Case,Nuk}.\\
An unexpected breakthrough came from two independent studies (in wildly different contexts) \cite{BD,BF03}: in \cite{BD} it was claimed that the average chord length is preserved even if the bounded medium is scattering (so the trajectories of entering particles are not straight lines but random paths), while the same result was stated in \cite{BF03}, when considering the average time spent by ants injected in a bounded domains before they leave it (see also \cite{anim}). This appears very surprising, since intuitively if we replace straight segments (or arcs) with erratic trajectories, we expect that a twofold mechanism deeply modifies the dynamics (stochastic short returns to the boundary, and long wandering walks in the domain without touching the boundary): see \cite{BF06}. A considerable effort has been conveyed in checking under which conditions this generalized Cauchy universality holds \cite{Maz04,Ben05,MMZ}: in particular the importance of homogeneous probability distribution inside the domain and of detailed balance in the scattering kernel have been pointed out. Remarkably,
Cauchy universality has  been recently supported by experimental results, in bacterial motion~\cite{Fran}, and light propagation in scattering media~\cite{Savo,Pier}. \\
On the other side, complex dynamics in bounded domains is sometimes associated with non homogeneous distributions, like in the case of active particles (see for instance \cite{Cap,VGV,YMM}), 
or in the presence of interactions with the walls~\cite{Sou})
so a natural question is whether Cauchy universality is maintained when nonuniform densities are present, or boundary effects are taken into account.\\
This is the main point we are going to discuss, providing examples where boundary conditions lead to nonuniform stationary probability distribution and violations of Cauchy universality; we will consider  both the case of rectilinear trajectories between successive collisions with the boundary (billiard) and the case where instead particles move in the interior of the body in a stochastic fashion (random walk).\\
In order to provide a simple geometric setting (which is typical of experimental realizations, too~\cite{Fran}) we will take $\Omega$ as the unit, 2-d, disk, and consider a particle moving with a constant speed ($v_0=1$) inside $\Omega$. To fix the dynamics we have to specify; ({\it i}) the way the particle moves inside $\Omega$ after a collision with the boundary and ({\it ii}) the rules prescribing the outgoing angle, once the ingoing angle is known, at a collision. Both prescription can be either deterministic or stochastic: the invariance of the average chord length (which in our units is $\tau_C =\pi/2$) has been verified up to now for both billiard and random walk case for elastic (specular reflection) boundary collisions~\cite{BD,BF03,Maz04,Ben05,MMZ}. While considering the distribution of random chords for a disk is a well defined procedure, recasting it in a billiard framework requires some care, since a circular billiard table with elastic reflections defines an integrable system. Before going on let us fix our notation: we will denote by $\mathbf{x}$ a generic point in the interior of $\Omega$ and by $\psi$ the angle between the speed of the particle and a fixed direction: in this way $(\mathbf{x}_t,\psi_t)$ are the coordinates of the point at time $t$ in the continuous time flow. When considering the collision map instead (mapping the position from one collision with the boundary to the next), the relevant coordinates will be denoted by $\mathbf{q}\in \partial \Omega$ and $\theta\in [-\pi/2,\pi/2]$ outgoing angle with respect to the normal vector $\mathbf{n_q}$ {(pointing inward) at $\mathbf{q}$ (alternatively we may use the the angle $\phi\in[0,2\pi]$, of the outgoing speed with respect to the oriented tangent at $\mathbf{q}$ ($\phi=\theta+\pi/2$).
In the case of a circular billiard table (with specular reflections), if we start a trajectory at $\mathbf{q}_0$ with outgoing angle $\theta_0$, chords will be all equal to $\tau_{\theta_0}=2 \cos(\theta_0)$ (in our units $R=1$). This is not due to an unfortunate choice of the billiard table, since all sufficiently smooth strictly convex billiard tables are non-ergodic~\cite{Laz} (for general references on billiard dynamics see \cite{Tab,ChMa,KH}). We can however average over the initial outgoing angle, by choosing an appropriate measure
\begin{equation}
\label{muchord}
\overline{\tau}=\int_{-\pi/2}^{\pi/2}\,\mu(d\theta_0)\tau_{\theta_0}.
\end{equation}
It is easy to check that we can recover Cauchy result ($\overline{\tau}=\tau_C=\pi/2$) if we choose $\mu(\theta_0)=\rho(\theta_0)d\theta_0=\frac 12 \cos(\theta_0)d\theta_0$. This is indeed the invariant distribution for the outgoing angle for a chaotic billiard with a uniform stationary distribution for the continuous flow (with specular reflections) \cite{ChMa,nC}; it also corresponds to the angular dependence of the flux generated by a uniform and isotropic distribution of particles entering the bounded region from outside. \\
An alternative derivation of this result (that provides the first example of our general setting) is to consider a random billiard, where the deterministic rule for the outgoing angle, at a collision point with the boundary, is replaced by a probability distribution density $\mathcal{P}_{\mathbf{q}}(\theta | \theta_{in})$ (which in principle may depend or not on the collision point $\mathbf{q}$ and the ingoing angle) 
. The Knudsen case~\cite{FY,CPSV,KCel} corresponds to the choice (Lambert reflections)
\begin{equation}
\label{kn}
\mathcal{P}_{\mathbf{q}}(\theta | \theta_{in})=\mathcal{P}_K(\theta)=\frac12 \cos(\theta).
\end{equation}
In this framework, checking for Cauchy universality consists in replacing rectilinear motion between collisions with a stochastic process: for instance in \cite{BF03,Ben05} a Pearson random walk is considered: the particle travels with a constant speed for a random time $t$ (or distance: since $v=1$) extracted from an exponential distribution
\begin{equation}
\label{dexp}
\mathcal{Q}_{E}(t)=\frac 1{\lambda} \exp(-t/\lambda).
\end{equation}
At the end of the rectilinear walk (at space point $\mathbf{x}$) the direction of the velocity is randomly reoriented, according to a chosen distribution density: in this paper (following \cite{BF03,Ben05} we will employ the random reorientation, the new direction $\psi_+$ is independent of the old orientation $\psi_-$: $\psi_+$ is a random variable uniform in $[0,2\pi]$ \cite{prep}. If the exponential excursion crosses $\partial \Omega$, it is reflected back according to the boundary conditions we are considering.\\
We first check the case of elastic boundary conditions (specular reflections): in this case we already know Cauchy universality holds, but this numerical experiment is useful to gauge how the asymptotic result is reached by increasing the statistics: see Fig. \ref{C-el}.
\begin{figure}
\includegraphics{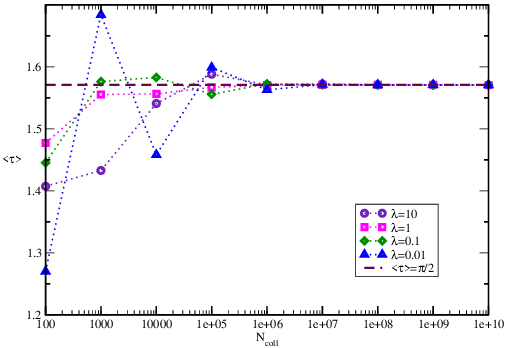}
\caption{\label{C-el}Time average $\langle \tau\rangle$ of the time between collisions, for increasing number of collisions, computed for different values of the parameter of the exponential distribution (\ref{dexp}) $\lambda$, and elastic reflections. The dashed line corresponds to $\tau_C$.}
\end{figure}
\\There are two physical motivations that suggest considering deviations from specular reflections: the case in which the agent is not properly modelled by a point particle (see for instance \cite{BG}), and possible roughness of the boundary~\cite{rou1,rou2}: other physical situations are modelled by non-standard boundary conditions as well~\cite{BCL}.
The first example we study is that of a fully random billiard~\cite{Evans}, where the particle moves ballistically between collisions, while the outgoing angle $\theta$ is a random variable, independent on the ingoing angle, uniformly distributed in $[-\pi/2,\pi/2]$. For this random billiard the dynamics enjoys strong ergodic properties~\cite{Evans,Fet}: we mention that such a billiard has also an independent interest in sampling problems of convex sets~\cite{Vem}. The average chord between collisions is easily computed as in (\ref{muchord}), where now $\mu_R(d\theta_0)=d\theta_0/\pi$, corresponding to a collision rule:
\begin{equation}
\label{rkn}
\mathcal{P}_{\mathbf{q}}(\theta | \theta_{in})=\mathcal{P}_R(\theta)=\frac1\pi,
\end{equation}
as a matter of fact:
\begin{equation}
\label{muchordR}
\overline{\tau}=\tau_R=\int_{-\pi/2}^{\pi/2}\,\mu_R(d\theta_0)\tau_{\theta_0}=\frac4{\pi}.
\end{equation}
This immediately shows that random billiards violate Cauchy universality: the key observation is that this is associated to a nonuniform stationary probability distribution in space, as we show now. We will denote the invariant probability density for the flow as
$\varrho(\mathbf{x},\phi)$ \cite{rot}: in all the cases we will consider rotational invariance is preserved (in biological settings however more complex patterns may arise~\cite{BB}), so
\begin{equation}
\label{r-dens}
\varrho (\mathbf{x},\phi)=\frac1{2\pi}g(r),
\end{equation}
where $r$ is the distance of $\mathbf{x}$ from the center of the disk $\Omega$.
In this way $\Phi(r)=2\pi r g(r)$ is the stationary probability distribution of the distance from the origin: for a uniform spatial probability distribution ($g(r)$ constant) we have $\Phi(r)=2r$.\\
Now consider a chord (of length $2 \cos (\theta_0)$) corresponding to an outgoing angle $\theta_0$: the values of $r$ along the chord range from $\sin(\theta_0)$ to one, and a uniform distribution along the chord leads to the corresponding $r$ density:
\begin{equation}
\label{cho-r}
\mathcal{W}_{\theta_0}(r)=\frac1{\cos(\theta_0)}\frac r {\sqrt{r^2-\sin^2(\theta_0)}}\quad  r\in[\sin(\theta_0),1].
\end{equation}
Now we evaluate the average of this expression, by using the appropriate measure, and by taking into account that the single $\theta_0$ contribution has to be weighted by the ratio of the chord length and the average $\tau_R$:
\begin{equation}
\label{wR}
{\wp}_R(\theta_0)=\frac{2 \cos({\theta_0})}{4/\pi};
\end{equation}
so
\begin{equation}
\label{RPr}
\Phi_R(r)=\int_{0}^{\arcsin(r)}\,\mu_R(d\theta_0){\wp}_R(\theta_0)\mathcal{W}_{\theta_0}(r).
\end{equation}
By changing variable $r\sin(\theta_0)=\sin(\alpha)$, we get
\begin{equation}
\label{RPr2}
\Phi_R(r)=r\int_0^{\pi/2}\,d\alpha\,\frac{1}{\sqrt{1-r^2\sin^2(\alpha)}}=rK(r^2),
\end{equation}
where $K(s)$ is the complete elliptic integral of the first kind~\cite{AbSt}. The expression Eq. (\ref{RPr2}) deviates from the linear behaviour corresponding to a uniform density (and it has a logarithmic singularity as $r\to 1$, due to the complete elliptic integral), so the random billiard both violates Cauchy universality and has a nonuniform spatial probability distribution (see Fig. \ref{Rdist}, first two lines), peaking close to the boundary. \\ Though our analytic argument in deriving Eq. (\ref{RPr2}) is not rigorous, the result can be validated by employ the findings in \cite{Evans}. \\
\begin{figure}
\includegraphics{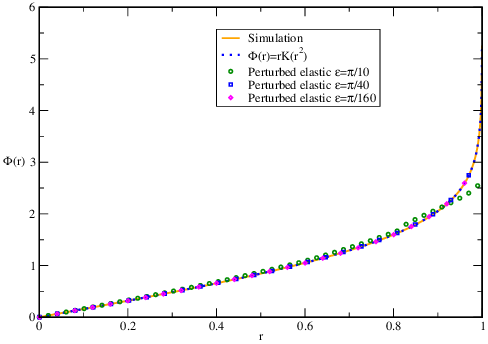}
\caption{\label{Rdist}Radial probability for the random billiard, and perturbed elastic reflections (see text). The numerical distributions were obtained by discretizing the trajectories with a $10^{-4}$ time step, considering $10^8$ collisions, and binning into $2500$ intervals the $r$ range.}
\end{figure}
As the random billiard leads to violation of Cauchy universality, a natural question is to check what happens when we substitute ballistic walks between collisions with an exponential random walk of parameter $\lambda$ (Eq. \ref{dexp}), with uniform direction 
resetting. This is illustrated in FIG. \ref{C-ra}. As we see there is no universal behaviour by varying $\lambda$, while when the average excursion grows, we are close to the random billiard estimate (\ref{muchordR}).
\begin{figure}
\includegraphics{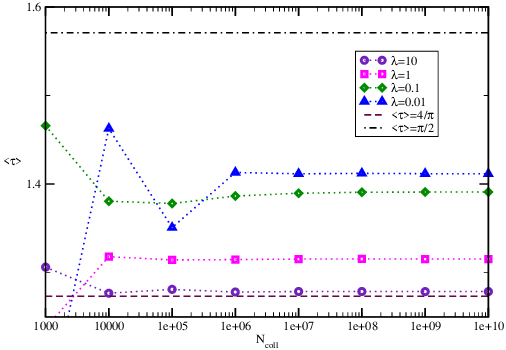}
\caption{\label{C-ra}Time average $\langle \tau\rangle$ of the time between collisions, for increasing number of collisions, computed for different values of the parameter of the exponential distribution (\ref{dexp}) $\lambda$, for the random billiard. The dashed line corresponds to $\tau_R$, while the dashed-dotted line is Cauchy value $\tau_C$.}
\end{figure}
\\Finally we consider another kind of boundary conditions, where possible roughness of the boundary is incorporated in a milder way than fully random reflections: the suggestion came from \cite{Mark}, where an $\varepsilon$ stochastic perturbation of elastic boundary conditions was shown to lead to ergodic behaviour for strictly convex billiard tables. More precisely, fix $\varepsilon\in [0, \pi/2]$, and denote by $\phi_{el}$ the outgoing angle (measured with respect to the oriented tangent), determined by the specular reflection law: the random elastic perturbed billiard is then defined by making the outgoing angle a random variable, uniformly distributed in the interval $[\phi_{el}-\varepsilon,\phi_{el}+\varepsilon]$, for $\phi_{el}>\varepsilon$ and $\phi_{el}<\pi-\varepsilon$: the rule must be modified when $\phi_{el}$ is sufficiently close to the tangent, to avoid orbits leaving the region $\Omega$. A possible choice is~\cite{Mark} to reset $\phi_{el}$ to $\varepsilon$ when $\phi_{el}< \varepsilon$, with the new outgoing angle uniformly distributed in $[0,2\varepsilon]$ (and the analogous prescription when $\phi_{el}$ is close to $\pi$). One expects that for very small $\varepsilon$ one should recover the universal behaviour, since the elastic reflection law is only slightly perturbed: numerical experiments however suggest that instead the behaviour is closer to the fully random billiard (see FIG. \ref{Rdist}): the radial distribution displays the same logarithmic (weak) singularity close to the boundary. 
\begin{figure}
\includegraphics{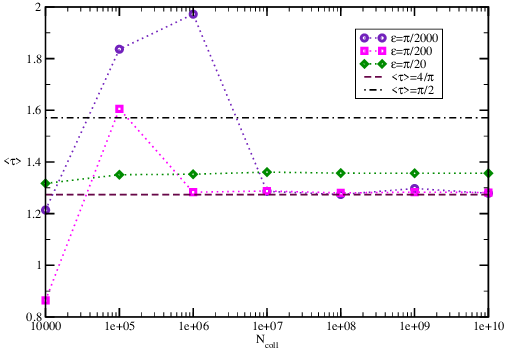}
\caption{\label{C-bM}Time average $\langle \tau\rangle$ of the time between collisions, for increasing number of collisions, computed for a billiard with random elastic perturbation (see text). For very small perturbations the results are very close to the fully random outgoing angle case. The dashed line corresponds to $\tau_R$, while the dashed dotted line corresponds to the Cauchy value $\tau_C$.}
\end{figure}
This somehow surprizing result is confirmed by simulations on the average chord, see FIG. \ref{C-bM}: when the perturbation is very small the simulations are very close to the random billiard value (\ref{muchordR}). A theoretical support for such findings comes from considering the reduced discrete dynamics for the outgoing angle $\phi_n \to \phi_{n+1}$: away from small intervals around $0$ and $\pi$, the stochastic dynamics is equivalent to a random walk with a uniform and symmetric jump distribution of width $\varepsilon$, so, away from the boundaries we expect a uniform stationary distribution~\cite{Frw}.\\
These features completely change when, while maintaining a stochastic perturbation of elastic reflections at the boundary, we turn from billiards to random walks (we still consider path segments generated by an exponential distribution of parameter $\lambda$ followed by a uniform random redirection of the velocity direction): a complete analysis is outside the scope of the present paper, we just point out when the mean free path $\lambda$ is sufficiently small, Cauchy universality is recovered (but the average chord does vary on increasing 
$\lambda$): see FIG. \ref{C-eM}
\begin{figure}
\includegraphics{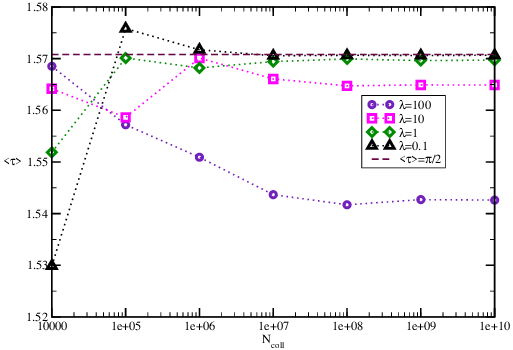}
\caption{\label{C-eM}Time average $\langle \tau\rangle$ of the time between collisions, for increasing number of collisions, computed for weakly stochastically perturbed elastic perturbation, $\varepsilon=\pi/200$ (see text), and an exponential random walk between collisions. For $\lambda$ not too big the results are very close to Cauchy estimate. The dashed line corresponds to $\tau_C$.}
\end{figure}
\\In conclusion, we have considered the average chord problem for either deterministic or stochastic motion in a bounded domain, for which remarkable universal properties have been proposed. Inspired by possible complex mechanism altering the specular reflection law for complex agents, we have considered random boundary conditions of different types: typically under these conditions Cauchy universality is violated, and this is associated to the appearance of nonuniform spatial densities. This may be relevant for other physically important problems of motion in confined systems, like the narrow escape problem \cite{Sch}, since enhanced spatial density close to the boundary increases the probability of hitting the target. Random reflections may also be relevant when considering reversible sticking to $\partial \Omega$ (see for instance \cite{Greb-imp,GrebK,Greb-imp2}), since when a particles sticks to the boundary, it is natural to consider the release outgoing angle as random, uncorrelated to the incoming direction.
\\
\begin{acknowledgments}
RA acknowledges an affiliation with INDAM. We thank Francesco Piazza and Francesco Ginelli for enlightening discussions at the early stage of this project.
\end{acknowledgments}


\begin{references}
\bibitem{rot} In all the cases we consider such a density is known to exist, see for instance \cite{FY,Evans,Fet}: moreover mixing is exponentially fast, which supports the reliability of our numerical simulations.
\bibitem{Frw} Actually the behaviour of the perturbation near the boundaries is crucial: while the form we are considering avoids sticking to $0$ or $\pi$: it is possible to choose other stochastic perturbations for which trajectories get captured at the boundary, see \cite{Mark}.
\end{references}
\end{document}